# Laboratory studies of THGEM-based WELL structures with resistive anodes

L. Arazi[*], M. Pitt[†], S. Bressler, L. Moleri, A. Rubin and A. Breskin

*Department of Particle Physics and Astrophysics*
*Weizmann Institute of Science, 76100 Rehovot, Israel*
E-mail: `lior.arazi@weizmann.ac.il`

ABSTRACT: In this work we investigate three variants of single amplification-stage detector elements; they comprise THGEM electrodes closed at their bottom with metallic or resistive anodes to form WELL-type configurations. We present the results of a comparative study of the Thick-WELL (THWELL), Resistive-WELL (RWELL) and Segmented Resistive WELL (SRWELL), focusing on their performance in terms of spark-quenching capability, gain variation with position and counting rate, pulse shapes and signal propagation to neighboring readout pads; the study included both $30 \times 30$ and $100 \times 100$ mm$^2$ detectors. It was shown that the WELL structures with resistive anodes offer stable operation even in a highly ionizing environment, with effective spark quenching, as well as higher gain than the standard THGEM/induction-gap configuration. Cross talk between neighboring readout pads was shown to be effectively eliminated in the segmented detector with a conductive grid underneath the resistive layer. The latter multiplier should allow for the design of very thin detectors, e.g. sampling elements in digital hadronic calorimeters planned for experiments in future linear colliders.

---

[*] Corresponding author
[†] Equal contributor

**Table of contents**

# 1   Introduction

Gas avalanche detectors share in one way or another some basic problems. Among them are counting-rate limitations due to charge evacuation from the avalanche region and gain limitations due to occasional discharges. The latter result from avalanche-induced secondary effects or from highly ionizing events: e.g. hadronic background when detecting MIPs, or MIPs when detecting single photoelectrons in RICH gaseous photon detectors. Unlike wire chambers, offering a relatively large dynamic range (due to charge saturation in the wire vicinity) but at the cost of limited rate capability, Micro-Pattern Gas Detectors (MPGDs) permit high-flux operation; however, they suffer from a relatively low dynamic range (no- or limited avalanche-charge saturation) – leading to occasional discharges. The subject is of high relevance and efforts have been made to limit discharges or reduce their energy – as to prevent damage to detector electrodes and to readout electronics, as well as to reduce after-spark dead-time. The reader is referred to [1]–[3] for more details on discharge mechanisms in gas-avalanche detectors.

    In this work, we concentrate on studies of several configurations based on the Thick Gas Electron Multiplier (THGEM) [4], [5], with particular focus on damping occasional radiation-induced discharges, aiming at the development of robust large-area detectors. The THGEM is

essentially a scaled-up (~ 10 fold) variant of the Gas Electron Multiplier (GEM) [6], suitable for applications requiring modest spatial resolutions. THGEMs are produced by standard printed-circuit board (PCB) technology: mechanical drilling of sub-millimeter diameter hole pattern through insulating (e.g. FR4) plates, copper-clad on both sides, followed by chemical etching of concentric insulating rims around the hole edges (the latter were found to considerably reduce the discharge probability, at the cost of some charging up effects [7], [8]). The thickness, hole diameter, pitch and rim size, may be chosen to meet the requirements of specific applications. A potential difference applied between the two faces of the THGEM electrode creates a strong dipole electric field within the holes; radiation-induced electrons generated in the drift region preceding the THGEM, or photoelectrons emitted from a photocathode deposited on the THGEM electrode [9], are focused into the holes, where they undergo charge multiplication. Very high gains can be reached in single- or cascaded-THGEM electrodes in a variety of gases. THGEM detectors offer sub-mm spatial resolution [10], few ns time resolution [11], robustness against occasional discharges [12], simple manufacturing and easy mechanical installation. The avalanche is confined within the holes, which act as independent multipliers with reduced photon-mediated secondary effects, even in poorly quenched and noble gases (e.g. Ne-mixtures with a few percent quencher which permit operation under very low voltages [7]).

In THGEM detector configurations studied extensively in previous works, the avalanche electrons induce detectable signals on a segmented readout anode during their drift towards it in an induction (collection) gap, with a typical width of 2 mm [7]. In this work we consider alternative structures, where the readout anode is in direct contact with the THGEM bottom electrode − closing the THGEM holes, in a so-called Thick WELL (THWELL) configuration. The closed-bottom geometry, also suggested in [13] and [14], is similar in its field shape to the WELL [15] and C.A.T. (the French acronym for "Compteur À Trou") [16]. Giving up the induction gap allows designing very thin detectors, which may be highly advantageous for applications where space limitations are important.

The new configurations studied here aim at reducing the energy and thus the consequences of occasional discharges. For that purpose, they incorporate thin resistive-film anodes - an idea that has been employed in the past in different detector configurations, e.g. see [10], [17], [18], and has been lately attracting much attention also in the MPGD community [19], [20]. The protection against discharges is two-fold: first, since the resistive layer is decoupled from the readout electrode, the readout electronics is not subject to direct high instantaneous currents during a discharge; second − as in Resistive Pate Chambers (RPCs) [21] - the relatively long clearance time of electrons from the resistive anode (here the bottom of the holes) leads to a substantial reduction of the local electric field and blocks the discharge before the entire charge on the detector is depleted.

One particular application which calls for a thin detector structure that will operate stably in a highly ionizing environment is in digital hadronic calorimetry (DHCAL), discussed for experiments in future colliders, such as the SiD experiment at ILC/CLIC [22], [23]. The baseline design of the SiD-DHCAL incorporates 40 layers of stainless steel absorber plates, interlaced with 8 mm thick gaseous sampling elements. These should have a lateral segmentation of 1 $cm^2$ readout pads, and should operate with high detection efficiency ($\gtrsim$ 95%) and close-to-unity pad multiplicity (number of pads activated per particle). The potential use of THGEM-based sampling elements for this application was discussed in previous publications [24]–[26], which included test-beam results of different THGEM configurations: single- and double-THGEM structures with an induction gap, a resistive WELL-THGEM (RWELL) and a Segmented Resistive-WELL (SRWELL) with a segmented thin-film resistive anode (the latter two with surface resistivity of 10-20 MΩ/square); the SRWELL was investigated both as a single-stage detector and in a cascaded configuration where it was preceded by a standard THGEM preamplifying element. The best results from the beam tests in terms of efficiency, pad-multiplicity and operation stability in hadronic environment were reached with the double-stage THGEM/SRWELL detector configuration [25], [26]; however single-stage configurations are still preferable in terms of cost and total detector thickness. The present work discusses the properties of the different single-stage THWELL structures in greater detail. Recent laboratory results of a new THGEM-based configuration, the Resistive-Plate WELL (RPWELL) - a THWELL electrode coupled to a thick resistive-plate anode of high bulk resistivity ($10^9 - 10^{12}$ Ωcm), are given elsewhere [27].

## 2 Experimental setup and methodology

### 2.1 Investigated THGEM-based structures

We investigated the following THGEM-based structures: Thick-WELL (THWELL), THWELL with a resistive anode (RWELL) and a segmented RWELL (SRWELL). The THGEM electrodes used in this work were $30 \times 30$ and $100 \times 100$ mm$^2$ in size, manufactured (Print Electronics Ltd, Rishon Lezion, Israel) by 0.5 mm diameter hole-drilling in 0.4 mm thick FR4 plates, copper-clad on one or two sides; the holes, with 0.1 mm wide etched rims, were arranged in an hexagonal lattice with a pitch of 1 mm in the double-faced electrodes, and in a square lattice with a pitch of 0.96 mm in the single-faced electrodes; the double-faced electrodes were used here as a reference to the standard THGEM configuration [4].

The THWELL is a single-faced THGEM electrode whose bottom is in direct contact with a readout metal anode, as shown in Figure 1a. The basic motivation for this structure was the reduction of the total thickness of the detector due to the absence of the induction gap. However, as shown below, the THWELL has an additional benefit, namely a higher gain for a given voltage and a higher maximum achievable gain, compared to the standard THGEM configuration (see section 3.1 below).

The RWELL (Figure 1b) is similar to the THWELL, but here the bottom side of the single-faced THGEM is closed by a resistive film deposited on a thin insulating sheet, with an underlying array of conductive readout pads; the latter pick up the signal inductively through the resistive anode. The resistive layers used in this work were produced by spraying a mixture of graphite particles and epoxy binder on 0.1 mm thick FR4 sheets [17]. The graphite concentration, the sheet thickness and the deposition conditions determine the surface resistivity of the layer; in this work the nominal resistivity was ~1-20 MΩ/square (higher resistivity values, although preferable for discharge quenching, are difficult to produce by this technique, as they involve large variations in local resistivity). The FR4 sheet with its resistive coating was mounted on top of the readout pads, immediately below and in direct contact with the bottom bare side of the single-faced THGEM electrode, forming the RWELL. A copper pad on the side of the coated FR4 sheet was used to connect the resistive layer to ground.

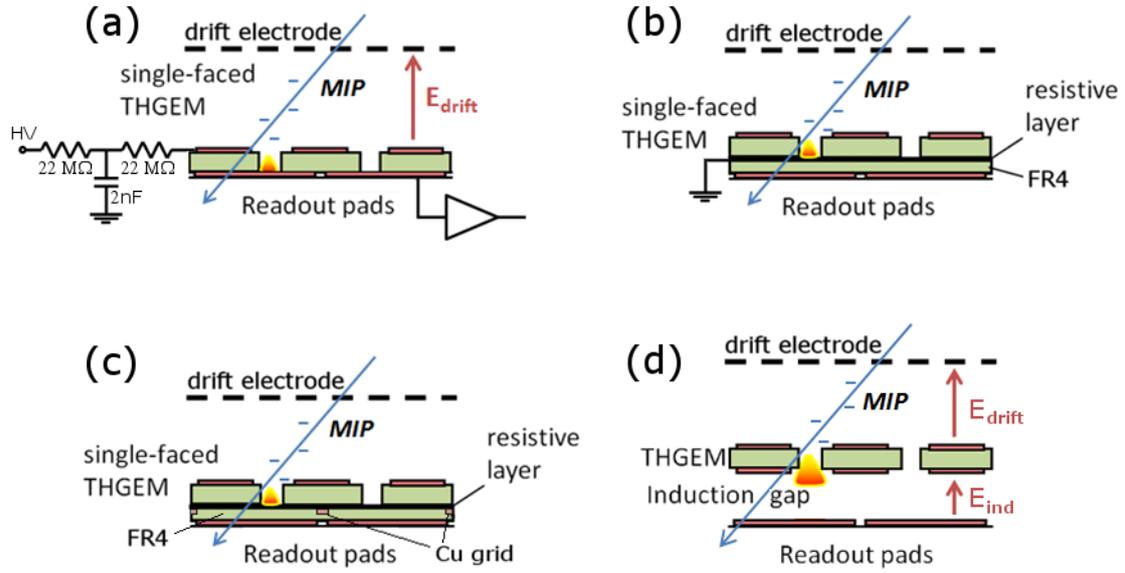

**Figure 1:** Schematic drawing of different THGEM detector configurations: THWELL (a), RWELL (b), SRWELL (c) and a THGEM with an induction gap (d). The detectors multiply charges deposited in a narrow drift gap. In the WELL configurations a single-faced THGEM electrode is set in direct contact with the anode: metal pads in the THWELL and a plain or segmented resistive film in front of the readout pads, in the RWELL and SRWELL, respectively. With metal anodes (as in the standard THGEM and THWELL configurations), the avalanche electrons are collected directly by the pads, while with resistive anodes the pads record the signals inductively. The THGEM top (and bottom, in (d)) and drift electrode were biased through an RC filter followed by an additional resistor as shown, for example in (a). The resistive layer (in (b) and (c)) was grounded, as shown in (b). The signal was taken, in all cases, from the readout pads, as shown in (a).

While the RWELL can effectively quench occasional discharges, it suffers from considerable pad cross-talk, resulting from the lateral diffusion of the avalanche electrons on the resistive layer across several neighboring readout strips or pads [24]. While this could be an advantage for analog avalanche center-of-gravity recording, in applications requiring segmented digital readout with minimal pad cross-talk (e.g. in the SiD-DHCAL) this would be a problem. The cross talk can dramatically reduced by adding charge-drain channels directly underneath the resistive film, in the so-called SRWELL configuration (Figure 1c), as demonstrated in [24, 25]. Such channels, namely a grid of thin conductive strips coinciding with the pad edges, allow for rapid clearance of the electrons diffusing on the resistive layer as they reach the pad boundary. In this study, the 0.1 mm thick FR4 sheet serving as the substrate for the resistive layer had a square grid of 0.1 mm wide Cu lines, 1 cm apart, matching the boundaries of the readout pads (Figure 2). The resistive layer was then sprayed on top, covering both the grid lines and the areas between them, with a surface resistivity of approximately 10-20 MΩ/square. This resistive anode was pressed against a segmented single-faced THGEM electrode, where the square lattice of holes covered only the pad area, with 1 mm wide bands (with no holes) matching the underlying resistive-anode grid lines. The purpose of the bands was to prevent avalanche formation above the Cu grid of the resistive layer, where spark-quenching is poor. The elements of the Segmented-RWELL (SRWELL) are shown in Figure 2.

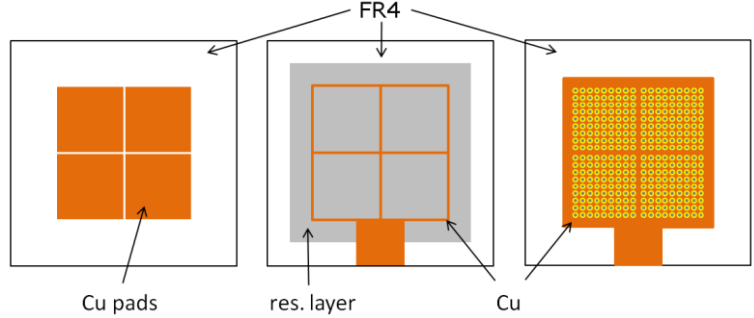

**Figure 2:** SRWELL-element components (here 2×2 pads): Left - the Cu pad array (connections not shown). Middle – the gridded resistive anode: the resistive film is deposited on top of a grid of thin copper lines printed on a thin FR4 sheet; the grid lines, matching the pads boundaries, rapidly drain avalanche electrons reaching the resistive layer, to reduce their diffusive spread to neighboring pads. Right – the segmented single-faced THGEM, with a square-hole pattern; the hole-less zones between the active THGEM ones, matching the grid line of the resistive anode, prevent avalanche formation in their vicinity (see text).

### 2.2 Gain measurements

*2.2.1 Gain curves*

Gain measurements were performed for the THWELL, RWELL (10 MΩ/square), SRWELL (10 MΩ/square) and double-faced THGEM with a 2 mm induction gap, using $30 \times 30$ mm$^2$ THGEM electrodes, mounted inside a 15 cm diameter aluminum chamber with a 50 μm thick Kapton window. The drift mesh used comprised 50 μm diameter stainless steel 304 wires arranged in a square pattern with 0.5 mm spacing. The drift gap was 6.5 mm in all cases. The detector was operated with Ne/CH$_4$(5%) at 1 atm with a nominal flow rate of 100 sccm. A two-stage oil bubbler was used at the outlet of the chamber. The detector was irradiated using an $^{55}$Fe 5.9 keV X-ray source at a rate of 1 Hz/mm$^2$, through a 5 mm diameter aperture in a 1.5 mm thick FR4 plate placed immediately in front of the drift mesh. The detector electrodes were biased using one or more CAEN N471A or N1471H power supply units, through low pass filters (R = 22 MΩ, C = 2 nF) and a 22 MΩ resistor in series (used to decouple the electrodes from the filter capacitor – see figure 1a). The drift field was 0.5 kV/cm in all cases. Measurements were done in pulse mode, with the 5.9 keV peak recorded using an amplification chain comprising a Canberra 2006 charge sensitive preamplifier, an Ortec 572A shaping amplifier and an Amptek MCA 8000A multi-channel analyzer. The amplification chain was calibrated before each gain-curve measurement using a pulse generator and 10 pF capacitor. The detector voltage was raised in 20 V steps until the appearance of the first spark within 5 minutes after changing the voltage; for comparative purposes the gain at this point is termed here the 'maximum achievable gain'.

*2.2.2 Gain homogeneity*

The production process of the single-faced 0.4 mm THGEM electrodes involves curing at high temperatures (typically about 180°C). This may cause some bending, potentially leading to gain variations across the electrode surface - if not perfectly pressed against the anode; this effect would come in addition to gain changes resulting from variations in the electrode thickness, as discussed in [8]. Small unwanted gaps between the THGEM and the anode would have two competing effects: on one hand, the avalanche will develop along a longer path, increasing the average effective gain; on the other, the larger gap between the top electrode and the anode will reduce the field within the THGEM holes, resulting in a reduced local gain. The standard double-faced large-area electrodes might also suffer from some thermal deformation; however, the avalanche path and the electric field within the holes remain practically unchanged for a sufficiently large (2 mm) induction gap. Comparing the gain homogeneity of standard and WELL-like structures can thus shed light on the potential effect of electrode deformation on the local gain.

Gain homogeneity was studied on $100 \times 100$ mm$^2$ single-faced detector electrodes mounted in THWELL, RWELL (10 MΩ/square) and SRWELL (20 MΩ/square) configurations and compared to a standard $100 \times 100$ mm$^2$ double-faced THGEM with a 2 mm induction gap. The detectors were mounted inside a 22 cm diameter aluminum vessel with a 200 μm thick, 20 cm diameter Mylar window. Measurements were done in Ne/CH$_4$(5%) at 1 atm with a typical flow of 40 sccm. The drift gap was 5 mm in all structures and drift field 0.5 kV/cm. Biasing and readout were done as in the gain measurements (here using an Ortec 124 charge sensitive preamplifier). The detectors were irradiated using a collimated (~3 mm spot diameter) X-ray $^{55}$Fe source, scanning an area of $60 \times 100$ mm$^2$, and the ratio between the local effective gain to its average value across the THGEM electrode was recorded.

### 2.2.3  *Rate dependence*

The slow clearance of avalanche electrons from the hole bottom of a WELL multiplier coupled to a resistive anode, may, in principle, be expected to reduce the detector gain at high rates (at a given applied voltage), as observed in RPCs [28] and recently in THGEM-based detectors coupled to resistive plate anodes with high bulk resistivity [27]. To determine whether this is in fact the case, the gain dependence on the irradiation rate was measured for all of the THGEM structures considered in this work.

The gain dependence on the detection rate was measured using an Oxford Instruments X-ray tube model XTF5011 with a copper anode operated at 20 kV, producing 8 keV X-rays superimposed on a low Bremsstrahlung profile. The measurements were performed on the $30 \times 30$ mm$^2$ electrodes mounted in the setup described in section 2.2.1 above, and comprised the RWELL, SRWELL, THWELL and standard THGEM + 2 mm induction gap structures. The RWELL was tested with both 1 and 10 MΩ/square resistive anodes, and the SRWELL with a 10 MΩ/square anode. The two *non-resistive* structures (THWELL and standard THGEM) were used here as a reference, to determine to what extent the variation of the gain with the detection rate indeed results from the presence of the resistive anode. The rate was controlled by varying the X-ray tube current and the number of 30 μm thick Cu filters attenuating the beam; it was increased monotonically, typically by a factor of 3-4 in each step, covering a range of ~0.1 Hz/mm$^2$ to ~$10^5$ Hz/mm$^2$ (the maximum rate was limited by the maximum allowable power of the tube and the geometry of the setup). The measurements were done in pulse-counting mode using the amplification and data acquisition chain described in section 2.2.1; the gain was determined by following the position of the 8 keV peak. In all cases the initial gain (at the lowest rate) was ~5000-8000: 660 V across the WELL-like structures and 760 V across the standard double-faced THGEM. The drift gap was 6.5 mm in all cases, and drift field 0.5 kV/cm. In each case, before starting the series of measurements the detector was irradiated overnight at the lowest rate. Care was taken to make sure that the gain is stable (i.e., does not change with time), by taking repeated spectra at each step. The time for gain stabilization was found to be of the order of 1 hour for the lowest rates, dropping to a few minutes at the highest rates. Up to ~$10^4$ Hz/mm$^2$ the beam spot on the detector was 5 mm in diameter (as defined by the aperture set in front of the drift mesh); for higher rates the beam spot size was limited to ~1 mm by additional collimators mounted on the X-ray tube; this was done in order to avoid pulse pileup.

**2.3 Pulse shapes**

*2.3.1 Rise-time*

The avalanche-induced pulse shape in the WELL-type configurations can be expected, based on Ramo's theorem [29], to be significantly different from that in the standard induction-gap structure. In the latter case, the anode is sensitive essentially only to the motion of avalanche electrons along the induction gap; the ion movement occurs within the THGEM hole, far from the anode, and has a negligible effect on the pulse shape. In contrast, in the WELL-type structures the ions movement occurs close to the anode, in a region that has a direct influence on pulse formation; this should manifest as a much longer rise time.

Anode pulse shapes were measured using an ORTEC 124 charge sensitive preamplifier and recorded by a digital oscilloscope (Tektronix TDS3052) with 50 Ω input impedance in AC coupling mode (with a frequency cutoff of 200 kHz). Pulses resulting from irradiation by an $^{55}$Fe source were recorded in the standard THGEM/induction gap configuration, as well as in the THWELL, RWELL (10 MΩ/square) and SRWELL (20 MΩ/square). This particular set of measurements was done with the 100×100 mm$^2$ detector, mounted in the setup described in section 2.2.2.

*2.3.2 Pulse shape on neighboring pads*

For applications requiring minimal pad cross-talk, good understanding of charge spreading across the resistive surface is essential. To study signal formation on neighboring pads in the RWELL and SRWELL configurations, we used an anode comprising 1×1 cm$^2$ pads behind the resistive layer. The measurements were performed on the 100×100 mm$^2$ electrodes. One pad ("primary") was irradiated through a 5 mm diameter hole, using an $^{55}$Fe source. Both the primary pad and its immediate neighbor ("secondary") were connected through charge sensitive preamplifiers (ORTEC 124) to a digital oscilloscope. By triggering on the primary-pad signal, both pulse shapes were recorded and compared.

A second experiment, aimed at measuring the propagation time of the diffusing electrons across the resistive layer and its dependence on the surface resistivity, was performed on a 30×30 mm$^2$ RWELL. The detector was operated, as before, with Ne/CH$_4$(5%) at 1 atm, here with a drift gap of 3 mm. Four nominal values of surface resistivity were investigated: 1, 2, 5, and 10 MΩ/square. The primary and secondary pads were connected through charge sensitive preamplifiers to a digital oscilloscope as described above. The X-ray beam (produced here using an Oxford Instruments generator model XTF5011) was collimated to ~0.5 mm diameter spot on the detector. The X-ray generator was mounted on a linear stage, allowing for accurate horizontal positioning relative to the pad boundary. Triggering was done on the primary-pad pulse and the time difference between the maxima of the primary- and secondary- pad pulses was recorded for varying distances of the beam from the pad boundary.

### 2.4 Discharges

#### 2.4.1 Discharge magnitude

The spark-quenching capabilities of the 100×100 mm$^2$ RWELL (10 MΩ/square) and SRWELL (20 MΩ/square) were studied quantitatively in comparison with the non-resistive THWELL and THGEM with induction gap configurations (see Figure 1), by observing the discharges and recording their magnitudes (here defined as the total charge released during a spark). Discharges were analyzed by reading the current delivered by the CAEN N471A power supply following a spark, using its $I_{MON}$ terminal. The $I_{MON}$ signal was recorded using a National Instruments data acquisition card (NI USB-6008). The total charge transferred from the THGEM top face to the anode upon the occurrence of a spark was calculated offline by integrating the $I_{MON}$ signal over time. Note that the total charge released during the spark might be composed not only of the charge stored on the detector electrodes, but also of that of the parasite capacitance of the coaxial cables and of the charge stored in the capacitor of the low-pass filter on the HV line connected to the THGEM top. To decouple the detector from the outer capacitance a 22 MΩ resistor was placed in series between the low-pass filter and THGEM (figure 1a). The coaxial cable between the THGEM top and 22 MΩ resistor was 20 cm long, adding a negligible parasitic capacitance to that of the THGEM itself. The charge distribution of an ensemble of ~30-200 discharges was recorded for each configuration, by irradiating the detector over 20-30 hours with 5.9 keV X-rays (here using a Philips PW 2215/20 X-ray generator), over a spot size of ~5 mm diameter on the THGEM surface. For the SRWELL the discharge magnitude (total spark discharge) was also measured for two beam positions: one at the pad center and the other at its boundary, where the discharge is developing close to the copper grid lines below the resistive film. The voltage provided by the power supply was also monitored and recorded (using the CAEN N1471H $V_{MON}$ terminal) for the SRWELL and THWELL, to observe their respective voltage drops during a spark.

## 3 Results

### 3.1 Gain measurements

Figure 3 shows the gain curves measured using 5.9 keV X-rays at rate of 1 Hz/mm$^2$ for the different WELL structures and for the standard 2 mm induction-gap configuration. The drift gap was kept at 6.5 mm, and the drift field was 0.5 kV/cm in all cases (as was the induction field in the standard configuration). The curves end at a THGEM voltage where the first spark occurs within the first five minutes of the measurement. The higher gain in the WELL structures results from the higher electric field inside the hole compared to that of the standard induction gap configuration (for the same voltage), as demonstrated in Figure 4 which shows the results of a calculation of the field along the hole axis (using Ansoft MAXWELL 3D v11); in WELL-type detectors the last generations of the multiplication avalanche are subject to a higher field than in the induction-gap structure, which leads to the observed higher gain. The higher value of the maximum achievable gain for these structures was also observed in [24]. The origin of this effect is not yet fully understood.

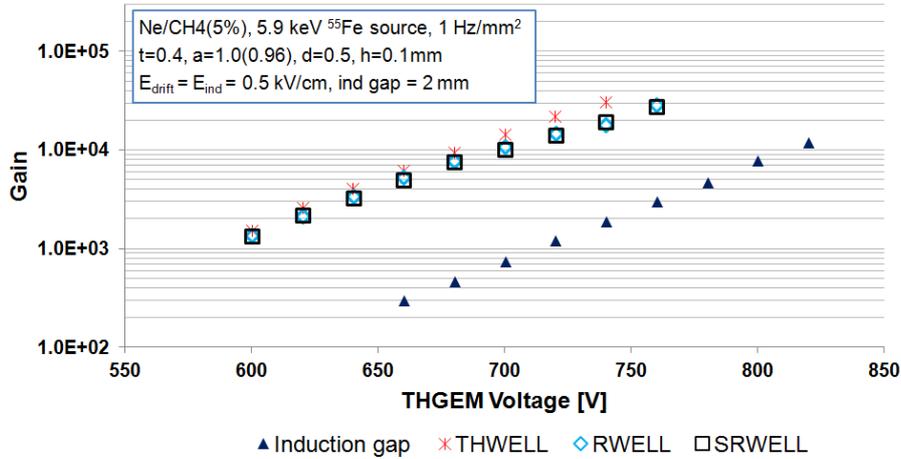

**Figure 3:** Gain curves for THWELL, RWELL (10 MΩ/square), SRWELL (20 MΩ/square) and standard THGEM with a 2 mm induction gap. The gains were measured in Ne/CH$_4$(5%) with 5.9 keV X-rays at a rate of 1 Hz/mm$^2$, over a 5 mm diameter spot size. Electrode parameters and experimental conditions are given in the figure.

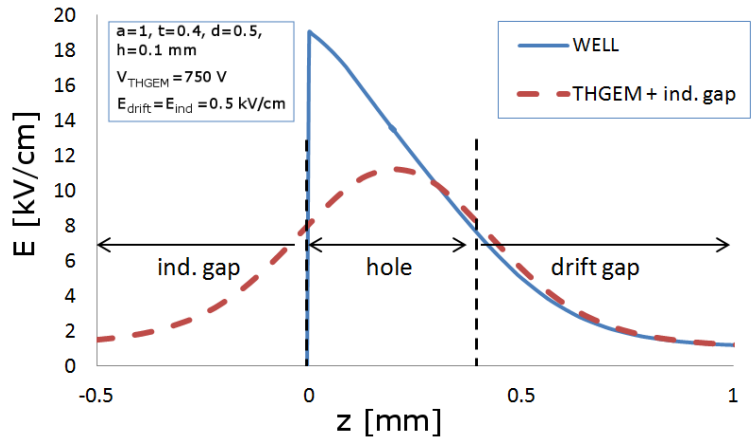

**Figure 4:** Comparison of the electric field inside the THGEM hole along the central axis in the induction-gap (Figure 1d) and THWELL (Figure 1a) configurations, as calculated using Ansoft MAXWELL3D v11. Electrode parameters and voltages are given in the figure.

The gain homogeneity measurements (local gain divided by its average value across the scanned area) are shown in Figure 5 for the different 100×100 mm$^2$ detector configurations. No specific trend in gain variation was observed. The standard deviation of the measured gain distribution was 7% for the THGEM/induction gap configuration and 9%, 16% and 13% for the THWELL, RWELL (10 MΩ/square) and SRWELL (20 MΩ/square) respectively. The deviations are somewhat higher for the WELL-type structures, possibly due to small imperfect contacts of the THGEM electrode and the respective anode.

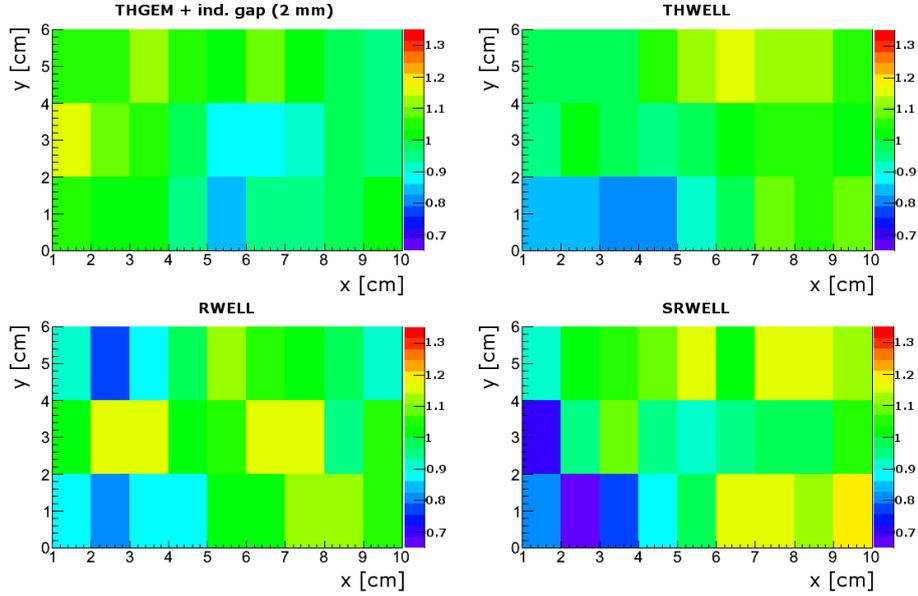

**Figure 5:** Gain variations (local gain divided by its mean value) across the 100×100 mm$^2$ detector electrodes for different configurations.

The gain dependence on the counting-rate, as measured for the THWELL, RWELL (with 1 and 10 MΩ/square resistive anodes), SRWELL (20 MΩ/square) and standard THGEM + 2 mm induction-gap configurations is shown in Figure 6. The WELL structures were all operated at a THGEM voltage of 660 V and the standard THGEM at 760 V. The drift field was 0.5 kV/cm in all cases (as was the induction field for the standard THGEM). The initial values of the gain for the investigated structures were: THWELL – 8100, RWELL (1 MΩ/square) – 5000, RWELL (10 MΩ/square) – 7800, SRWELL (20 MΩ/square) – 6100, standard THGEM – 5200. The curves were all arbitrarily normalized to unity at a rate of 0.3 Hz/mm$^2$. Several repetitions were made for the THWELL and RWELL, showing a spread of the measured values comparable in size to the variance across the different WELL structures, evident in Figure 6. All WELL-type structures showed a similar behavior, namely a monotonic decrease of the gain with the rate, with no initial plateau and approximately the same slope; in all cases, the gain drops by ~50% from its value at ~0.2 Hz/mm$^2$ at rates of ~5·10$^4$ Hz/mm$^2$. The standard THGEM/induction gap structure, on the other hand, showed about a two-fold smaller gain-drop over the investigated range of rates, consistent with previous measurements [3]. The fact that the non-resistive THWELL shows the same behavior as the resistive structures indicates that the underlying effect is not related to the presence of the resistive layer. It is rather likely related to charging up of the hole's wall. The possibility of charging up of the upper rim surrounding the hole is not consistent with the marked difference between the THWELL and standard THGEM, which share the same geometry in the top part of the hole. As shown below, the typical time scale for electron diffusion on the resistive layer out of the hole bottom should be of the order of 1 µs – similar to the clearance time of avalanche ions. Thus, the effect of electron clearance should be evident at rates close to ~$10^6$ Hz/mm² – one order of magnitude higher than the investigated range of rates.

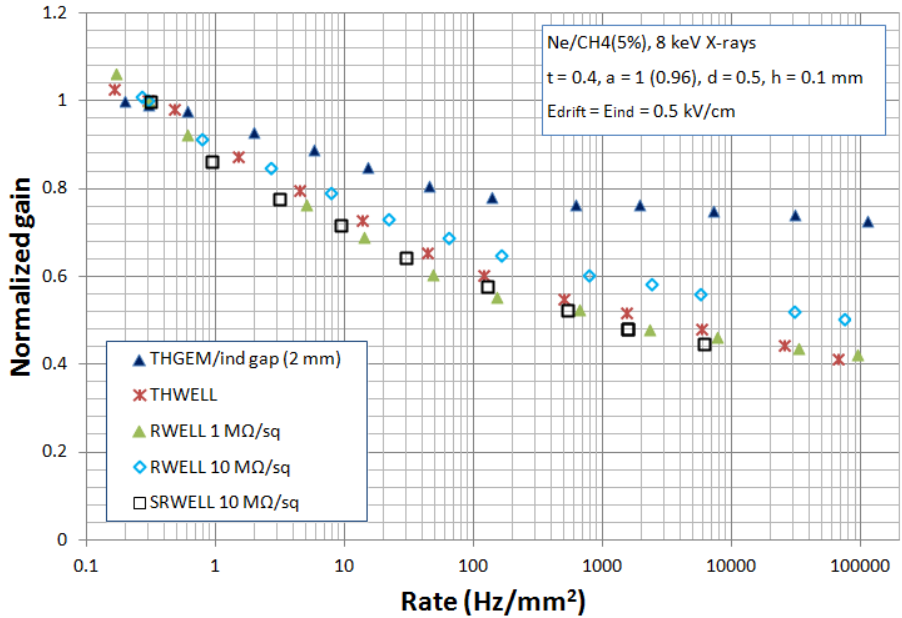

**Figure 6:** Rate dependence of the gain in the THWELL, RWELL (1 and 10 MΩ/square anodes), SRWELL (20 MΩ/square) and standard THGEM + 2 mm induction gap configurations. Electrode parameters and experimental conditions are given in the figure. The curves are normalized to unity at 0.3 Hz/mm². The initial gains were between 5000 and 8100 (see text for details).

Note that the methodology employed in this work to measure the gain dependence on the detection rate is different than the one employed previously in [24]. In that work, the gain of an RWELL detector was inferred, for rates in the range $1 \cdot 10^4 - 2 \cdot 10^6$ Hz/mm², from changes in the dc current of the anode, rather than from changes in the 8 keV peak position as was done here. The gain values shown in [24] were displayed on a logarithmic scale, rather than on a linear one as in Figure 6, leading to an impression of a plateau in the range $10^4 - 10^5$ Hz/mm²; however, when plotted on a linear scale, both data sets show a consistent relative gain drop of 10% over this range. Recent measurements [27] of the gain dependence on the detection rate in WELL-type configurations (THWELL, RWELL and RPWELL), were performed using a similar method as the one employed here, but with several differences: shorter stabilization times on each step, 0.8 mm thick electrodes (compared to 0.4 mm here) and much smaller spot size (~1 mm compared to ~5 mm here); the results in [27] are thus somewhat different then those reported here, in particular for the THWELL, where a smaller relative gain drop in lower rates was observed (the RWELL curve in [27] is similar to those shown in Figure 6)).

## 3.2 Pulse shapes

Typical pulse shapes of the THWELL, RWELL and SRWELL are shown in Figure 7, in comparison with that of the standard THGEM/induction gap configuration. The ion component is visible for all WELL-based structures, with a ~1 µs rise time, while for the standard THGEM configuration only the fast electron component is observed. Note that the pulse shapes of the WELL-type structures also begin with a rapid rise, resulting from the movement of avalanche electrons towards the anode.

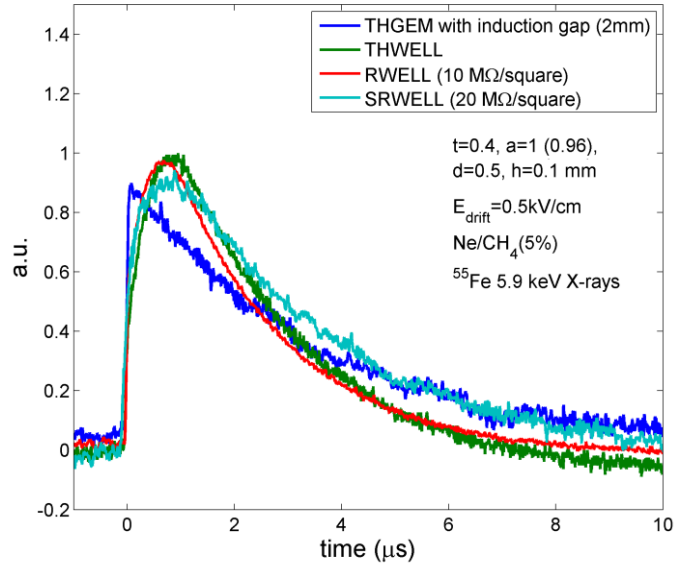

**Figure 7:** Typical pulse shapes of the different WELL structures and of an induction-gap configuration (shown in Figure 1) measured with a charge sensitive preamplifier.

Figure 8 shows typical pulse shapes recorded on the primary pad and its immediate neighbor ("secondary pad"), for the RWELL (10 MΩ/square) and SRWELL (20 MΩ/square), with the beam set to irradiate the primary pad's center. In the RWELL, the neighbor pad picks up a delayed signal, with a long rise-time of a few µs and amplitude comparable to that of the primary pad. In contrast, in the SRWELL the secondary pad picks up only a negative low-amplitude signal resulting from capacitive coupling between the two pads (the same feature is also observed in the RWELL case). Note that the long rise time of the secondary pad pulse in the RWELL case can be used, in principle, to discriminate it from that of the primary pad and thus reduce the pad multiplicity.

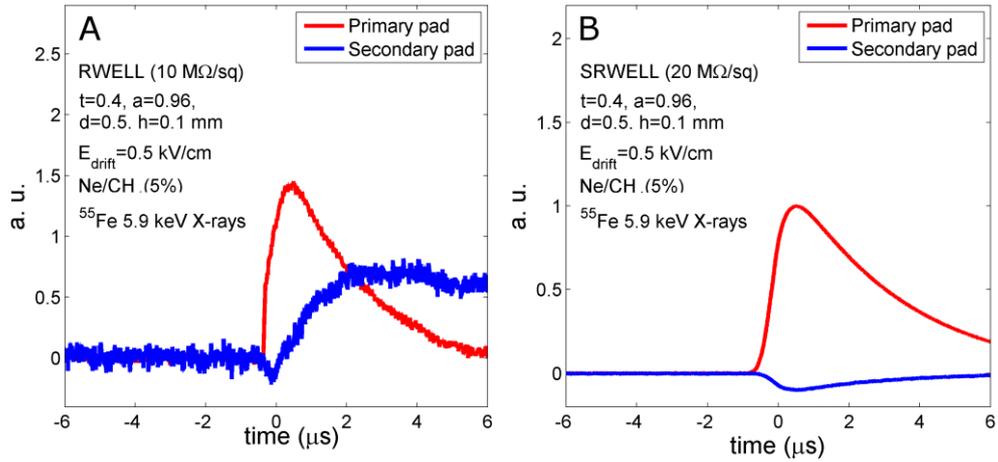

**Figure 8:** Typical x-ray induced pulse shapes measured from the primary and secondary pads, for RWELL (A) and SRWELL (B); in both cases the beam was set to irradiate the primary pad's center (5 mm from its boundary). Electrode parameters and experimental conditions are given in the figure.

The delayed rise of the pulse on the RWELL secondary pad results from the slow propagation of electrons on the resistive layer and can be expected to increase with its surface resistivity. This effect is shown in Figure 9, for a 30 ×30 mm² RWELL with varying surface resistivities. The graph shows the effective signal-propagation velocity $\delta x/\delta t$, where $\delta x$ is the distance of the beam from the pad boundary, and $\delta t$ is the time difference between the maxima of the primary and secondary pad pulses.

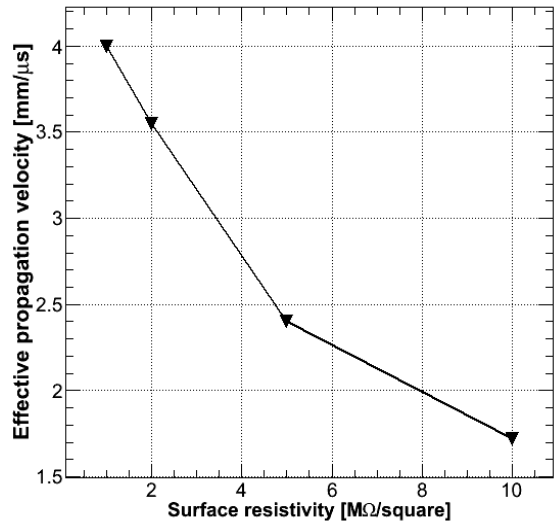

**Figure 9**: Effective signal propagation velocity across the RWELL resistive anode determined, for different resistivity values, from the distance of the beam from the boundary and from the time interval between the maxima of the primary and secondary pads' pulses.

## 3.3 Discharges

The discharge magnitude was measured for the 100 ×100 mm$^2$ 10 MΩ/square RWELL, 20 MΩ/square SRWELL, THWELL and standard THGEM configuration with a 2 mm induction gap and 0.5 kV/cm induction field, using 8 keV X-rays irradiating a spot of ~5 mm diameter. For the SRWELL, measurements of discharge magnitudes were also performed with the X-ray beam irradiating either the pad center or the 'blind' copper strips above the grid lines (with the same 5 mm spot size). The resulting discharge-magnitude distributions are shown in Figure 10. The standard THEGM with induction gap and THWELL configurations, show similar narrow distributions with the discharge-induced charge of the order of the total charge stored on the detector (~400 nC, corresponding to a capacitance of a few hundred pF); in some cases the sparks have double or triple the charge, probably resulting from consecutive discharges occurring on a short time scale. In contrast, both the RWELL and SRWELL show considerable spark-quenching capabilities, with a ~10-fold and ~5-fold reduction of the spark charge, respectively (with respect to the average values of the distributions). The reduced spark-quenching factor of the SRWELL is likely the result of the much shorter distance the electrons diffusing on the resistive layer need to cross to get to ground potential (leading to a smaller effective resistance along their path). Note that for the SRWELL, no noticeable difference was observed between discharges developing close to the center pad and those occurring near the grid lines.

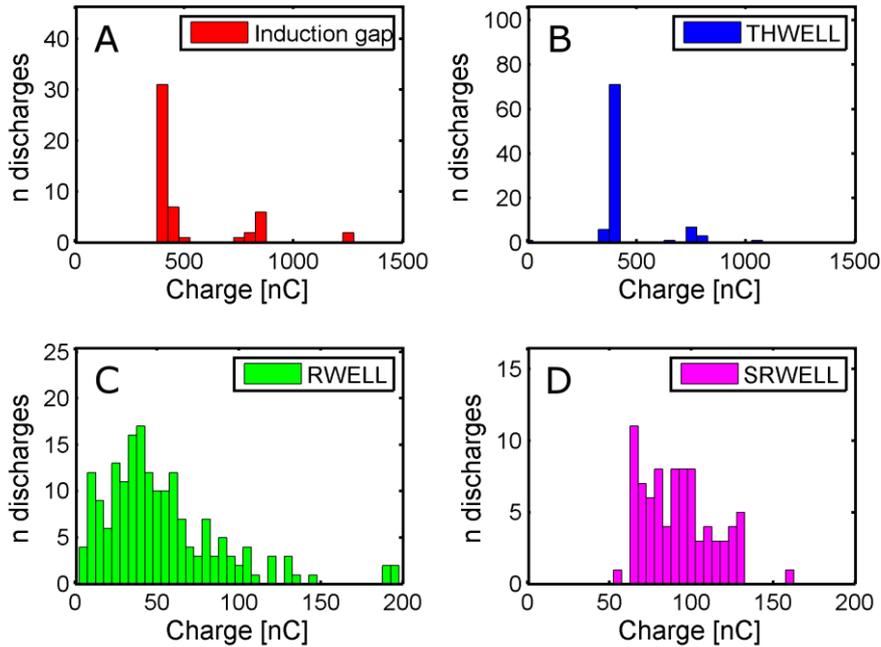

**Figure 10:** Discharge magnitude histograms of the different configurations (A-D). See text for experimental parameters.

In spite of the somewhat lower spark quenching capability of the SRWELL compared to the RWELL, it still shows stable operation even during high ionizing events. Figure 11 shows a comparison between the power supply voltage and current output (V$_{MON}$ and I$_{MON}$), supplied to the THGEM top electrodes, for the THWELL (left) and SRWELL (right). SRWELL sparks are characterized by a significantly smaller associated currents and essentially no voltage drops.

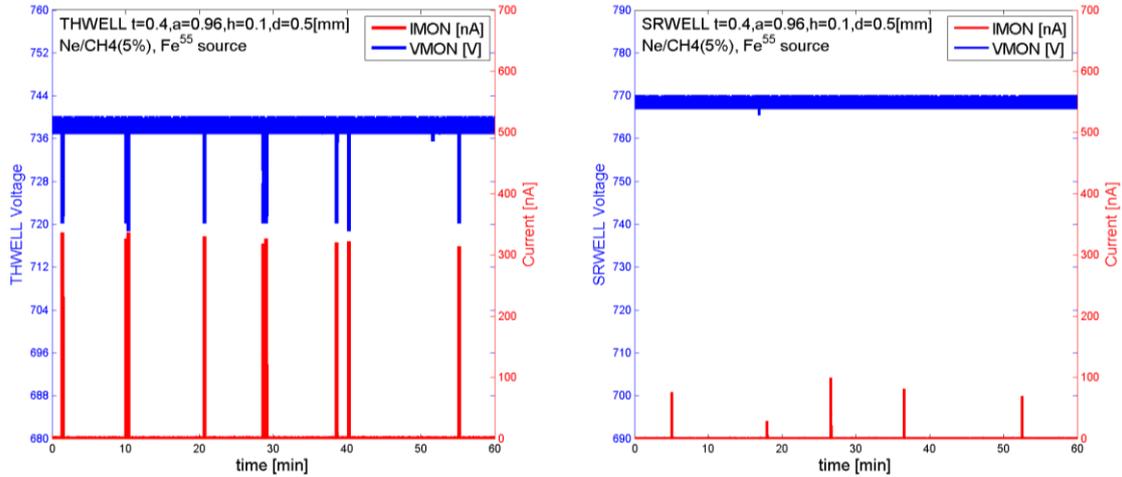

**Figure 11:** Comparison of the voltage and current monitored from the power supply, for the THWELL (left) and SRWELL (right) configurations; both measurements lasted 1 hour at a gain slightly higher than the nominal maximum achievable gain ($\sim 3 \times 10^4$). Electrode parameters and experimental conditions are given in the figure.

## 4 Summary and discussion

This work focused on investigating the key properties of three THGEM-based WELL-type configurations: the THWELL, RWELL and SRWELL. Starting with the inherent advantages of the THGEM, namely its robustness and scalability to large detection areas, these new structures are part of our effort to develop thin, spark-resistant large dynamic range detectors, suitable for applications such as (but not limited to) the digital hadronic calorimeter (DHCAL) planned for the SiD experiment of the ILC/CLIC.

The simplest structure, the THWELL, besides being thinner than a standard THGEM/induction gap configuration, also displayed a 10-fold higher gain at the same voltage, due to the larger field close to the hole bottom, and a larger maximum achievable gain. This higher gain allows MIP detection in a single amplification stage. A potential disadvantage of the THWELL, however, is that when a discharge occurs, its entire charge flows directly to the readout electronics with possible harmful consequences. One approach to mitigate this can be to add protective elements directly into the readout circuit itself; another option, employed here in the RWELL structure, is to place a continuous resistive layer, deposited on top a thin insulating sheet, between the single-faced THGEM and readout pads. The resistive layer has two advantages: first, it protects the readout electronics from direct discharge currents; second, it reduces the amount of charge flowing during a discharge by a large factor ($\sim 10$ in this work for $100 \times 100$ mm$^2$ detector). The use of a continuous resistive layer, however, has one disadvantage compared to the THWELL, namely that unless special signal processing algorithms are employed, a single avalanche results in multiple pad triggering because of the diffusive spread of electrons across the resistive surface. For application such as DHCAL, where low pad multiplicity is essential, modifying the RWELL to the segmented-RWELL (SRWELL) solves this problem. The SRWELL with a resistive layer having a square grid of thin copper lines underneath, matching the pad boundaries, displayed negligible cross-talk between adjacent pads (also demonstrated in [25,26]), while still preserving a significant spark-quenching capability.

A common feature of all of the WELL-type structures studied here is the monotonic decrease of the gain with the detection rate - an effect which is considerably more pronounced in these structures than in the standard THGEM/induction gap configuration; quantitatively, this gain drop amounts to ~50% at a rate of ~$10^5$ Hz/mm$^2$. As noted in the text, the observation that this behavior appears not only in the RWELL and SRWELL, but also in the *non-resistive* THWELL, suggests that the underlying mechanism is not related to electron clearance from the hole bottom by lateral diffusion across the resistive layer. This is in contrast to recent observations made on WELL structures coupled to a resistive plate (RPWELL), where the gain drop increased with the bulk resistivity of the anode [27]. The difference between these two observations can be attributed to the much longer characteristic time of electron transport through the resistive plates – of the order of $10^{-4} - 10^{-1}$ s for bulk resistivities in the studied range of $10^9 - 10^{12}$ Ωcm - compared to a typical time scale of $10^{-6}$ s observed in this work. The gain drop in the WELL structures studied here may be explained by charging up of the hole wall (a common feature to the THWELL, RWELL and SRWELL). A complete explanation, however, requires further studies and is beyond the scope of this work.

While the observed gain drop with rate is non negligible, it may not pose a problem in applications involving a constant average flux of particles, as long as the detector is stable (at the set gain) in the presence of high ionization background (especially if the rate is considerably smaller than $10^5$ Hz/mm$^2$). If the rate does change with time, the gain drop may still not be a problem if one is interested in digital readout only, as in purely digital hadronic calorimetry.

One might have expected that the "blind" hole-less bands included in the SRWELL design would result in reduced detection efficiency for events occurring close to the pad boundary. This effect, however, was shown to be small (~2%) in a beam study with muons and pions (as discussed in [25], [26]). Another potential concern is the formation of small gaps between the WELL electrodes and the resistive layer, due to the possible non-planarity of the former. In the present study this had no observable effect on the measured local gain. For larger electrodes, this potential problem can be effectively mitigated, for example by adding small fixations at selected points across the electrode.

To conclude, the WELL-type THGEM-based detectors investigated in this work have several advantages – small thickness, higher gain, effective spark quenching (for the RWELL and SRWELL) and negligible cross-talk between neighboring pads (for the SRWELL detector). While the gradual drop in gain with the counting rate may pose a problem in some high-rate applications, it is probably not prohibitive when digital particle counting is required (as in the SiD-DHCAL) – provided the gain can be made large enough to ensure high detection efficiency even when the gain drops.

## Acknowledgements

This work was largely motivated by DHCAL R&D. We would like to thank Prof. J. F. C. A. Veloso and Dr. C. D. R. Azevedo from Aveiro University, Prof. J. M. F. dos Santos and Dr. H. Natal da Luz from the University of Coimbra, and Prof. A. P. White from the University of Texas at Arlington for their interest and fruitful discussions regarding this work. We would further like to thank Dr. S. Shilstein from the Weizmann Institute for his help in constructing the X-ray irradiation setup used in this study, and Mr. S. Cohen and Mrs. L. Gaffry for their help in the measurements. This work was supported in part by the Israel-USA Binational Science Foundation (Grant 2008246) and by the Benozyio Foundation. A. Breskin is the W.P. Reuther Professor of Research in the Peaceful use of Atomic Energy.